\documentstyle[12pt]{article}
\newcommand{\be}{\begin{eqnarray}}
\newcommand{\ee}{\end{eqnarray}}
\newcommand{\non}{\nonumber}

\begin{document}

\begin{titlepage}
\strut\hfill UMTG--199
\vspace{.5in}
\begin{center}

\LARGE Boundary $S$ Matrix for the XXZ Chain \\[1.0in]
\large Anastasia Doikou, Luca Mezincescu and Rafael I. Nepomechie\\[0.8in]
\large Physics Department, P.O. Box 248046, University of Miami\\[0.2in]  
\large Coral Gables, FL 33124 USA\\

\end{center}

\vspace{.5in}

\begin{abstract}
We compute by means of the Bethe Ansatz the boundary $S$ matrix for 
the open anisotropic spin-${1 \over 2}$ chain with diagonal 
boundary magnetic fields in the noncritical regime ($\Delta > 1$).
Our result, which is formulated in terms of $q$-gamma functions, 
agrees with the vertex-operator result of Jimbo {\it et al.}
\end{abstract}
\end{titlepage}

\section{Introduction}

The concept of boundary $S$ matrix in $1+1$ dimensional integrable 
quantum field theory was precisely formulated by Ghoshal and 
Zamolodchikov in Ref.  \cite{GZ}.  They also developed there a 
bootstrap approach for determining such $S$ matrices.  Boundary $S$ 
matrices can also be computed for integrable quantum spin chains by a 
direct Bethe-Ansatz approach which was proposed in Ref.  \cite{GMN}.
\footnote{This method is a generalization of the approach developed by 
Korepin-Andrei-Destri \cite{korepin}, \cite{andrei/destri} to 
calculate bulk two-particle $S$ matrices. See also \cite{FS}.} 
This method has been used until now only for isotropic models 
\cite{GMN}, \cite{essler} - \cite{tsuchiya}.  We recently simplified 
this method in Ref.  \cite{DMN}.  In the present Letter, we take 
advantage of this simplification to analyze an {\it anisotropic} 
model, namely the open XXZ spin chain
\be
{\cal H} =  {1\over 4}\Big \{ \sum_{n=1}^{N-1} \left( 
\sigma^{x}_n \sigma^{x}_{n+1}
+ \sigma^{y}_n \sigma^{y}_{n+1} 
+ \cosh \eta \ \sigma^{z}_n \sigma^{z}_{n+1} \right)  \non \\
+ \sinh \eta \coth (\eta \xi_-) \ \sigma^z_1 
+ \sinh \eta \coth (\eta \xi_+) \ \sigma^z_N 
\Big\} \,, \label{open} 
\ee 
where the real parameters $\xi_\pm > 1/2$ correspond to boundary magnetic 
fields.  For simplicity, we restrict our attention to the case $\Delta 
\equiv \cosh \eta > 1$, which corresponds to the noncritical regime in 
which there is a nonzero gap.  (See e.g.  Ref.  \cite{JM}.) Our result 
for the boundary $S$ matrix, which is formulated in terms of $q$-gamma 
functions \cite{GR}, agrees with the result found by Jimbo {\it et al.} 
\cite{jimbo} by means of the vertex operator approach.  In the limit 
$\eta \rightarrow 0$, we recover the results of Refs.  \cite{GMN} and 
\cite{DMN}.

%We now outline the contents of the paper. In Sec. 2, we review the 

\section{Bethe Ansatz and one-hole state}

In this Section we review the exact Bethe Ansatz (BA) solution of the 
open XXZ chain, and we compute the root and hole density for the BA 
state consisting of a single hole.
  
The eigenvalues of ${\cal H}$ and 
$S^z = {1\over 2}\sum_{n=1}^{N} \sigma_{n}^{z}$ are given by
\cite{gaudin1} - \cite{sklyanin} 
\be
E &=& - {1\over 2} \sinh^{2} \eta \sum_{\alpha =1}^{M} {1\over 
\sin  \eta \left( \lambda_{\alpha}  - {i\over 2} \right) 
\sin  \eta \left( \lambda_{\alpha}  + {i\over 2} \right) }  
\,, \label{energy} \\ 
S^{z} &=& {N\over 2} - M \,, \label{spin}
\ee 
where $\lambda_{1} \,, \cdots \,, \lambda_{M}$ satisfy the BA
equations
\be
{\sin  \eta \left( \lambda_{\alpha} + i( \xi_{+} - {1\over 2}) 
\right) \over 
 \sin  \eta \left( \lambda_{\alpha} - i( \xi_{+} - {1\over 2}) 
\right) }
{\sin  \eta \left( \lambda_{\alpha} + i( \xi_{-} - {1\over 2}) 
\right) \over 
 \sin  \eta \left( \lambda_{\alpha} - i( \xi_{-} - {1\over 2}) 
\right) }
\left( {\sin  \eta \left( \lambda_{\alpha} + {i\over 2} \right) 
\over   \sin  \eta \left( \lambda_{\alpha} - {i\over 2} \right) }
      \right)^{2N} \non \\
= \prod_{\scriptstyle{\beta=1}\atop \scriptstyle{\beta \ne \alpha}}^M 
{\sin  \eta \left( \lambda_{\alpha} - \lambda_{\beta} + i \right) 
\over 
 \sin  \eta \left( \lambda_{\alpha} - \lambda_{\beta} - i \right) }
{\sin  \eta \left( \lambda_{\alpha} + \lambda_{\beta} + i \right) 
\over 
 \sin  \eta \left( \lambda_{\alpha} + \lambda_{\beta} - i \right) }
\,, \qquad \alpha = 1 \,, \cdots \,, M \,. 
\label{BA}
\ee
(We neglect in Eq. (\ref{energy}) additional terms which are independent 
of $\{ \lambda_{\alpha} \}$.)

Introducing the notation
\be
e_n(\lambda) = 
{\sin  \eta \left( \lambda + {in\over 2} \right) 
\over           
 \sin  \eta \left( \lambda - {in\over 2} \right) } \,,
\qquad
g_n(\lambda) = 
{\cos  \eta \left( \lambda + {in\over 2} \right) 
\over           
 \cos  \eta \left( \lambda - {in\over 2} \right) } \,,
\ee
the BA equations take the more compact form
\be
e_{2\xi_{+}-1}(\lambda_{\alpha})\ e_{2\xi_{-}-1}(\lambda_{\alpha})\ 
g_{1}(\lambda_{\alpha})\  e_{1}(\lambda_{\alpha})^{2N+1} 
= - \prod_{\beta = 1}^{M} e_{2}(\lambda_{\alpha} - \lambda_{\beta})\
e_{2}(\lambda_{\alpha} + \lambda_{\beta}) \,.
\label{compact}
\ee
Note that the factor $g_{1}(\lambda_{\alpha})$ is absent in the 
isotropic limit $\eta \rightarrow 0$.

Without loss of generality, we restrict $\eta > 0$. Moreover, the 
requirement that BA solutions correspond to independent
BA states leads to the restriction  (see \cite{GMN} and 
references therein)
\be
Re \left( \lambda_{\alpha} \right) \in \left[ 0 \,, {\pi\over 2 \eta} 
\right] \,, \qquad 
\lambda_{\alpha} \ne 0 \,, {\pi\over 2 \eta} \,.
\ee

Following \cite{DMN}, we focus now on the BA state consisting of a 
single hole.  This state lies in the sector $N=$ odd with $M = {N\over 
2} - {1\over 2}$ and $\{ \lambda_\alpha \}$ real.  This state has $S^z 
= +{1\over 2}$.

Since Eq. (\ref{compact}) involves only products of phases, 
it is useful to take the logarithm. In this way we arrive at the 
desired form of the BA equations
\be 
h (\lambda_\alpha) = J_\alpha \,, 
\label{BAlog}
\ee 
where the so-called counting function $h(\lambda)$ is given by
\be 
h(\lambda) &=& {1\over 2\pi} \Big\{ (2N +1)q_1(\lambda) + r_1(\lambda) 
+ q_{2\xi_+ - 1} (\lambda) + q_{2\xi_- - 1} (\lambda)  \non \\ 
& &  - \sum_{\beta=1}^{M} \left[ q_{2}(\lambda - \lambda_{\beta}) 
+ q_{2}(\lambda + \lambda_{\beta}) \right] \Big\}
\,, \label{counting} 
\ee 
$q_n (\lambda)$ and $r_n (\lambda)$ are odd monotonic-increasing functions 
defined by 
\be
q_n (\lambda) &=& \pi + i\log e_n(\lambda)
\,, \qquad -\pi < q_n (\lambda) \le \pi \,, \\ 
r_n (\lambda) &=&  i\log g_n(\lambda)
\,, \qquad \quad -\pi < r_n (\lambda) \le \pi \,,
\label{qr}
\ee
and $\{ J_\alpha \}$ are certain integers which serve as ``quantum numbers'' 
that parametrize the BA state. (See e.g. \cite{FT}.)

We shall need in the next section the root and hole density 
$\sigma(\lambda)$ for the one-hole BA state, which is defined by
\be
\sigma(\lambda) = {1\over N} {d h(\lambda)\over d \lambda} \,.
\label{sigma1}
\ee 
To calculate this quantity, we must pass with care from the sum in 
$h(\lambda)$ to an integral. Indeed, with the help of the 
Euler-Maclaurin formula for approximating sums by integrals, and 
using the fact that the solutions $\lambda = 0 \,, {\pi\over 2\eta}$ 
of the BA equations are excluded, one can derive\footnote{The argument 
is a slight modification of the one presented in the appendix of 
Ref. \cite{GMN}.} for a state with $\nu$ holes the following general result:
\be 
{1\over N}\sum_{\alpha=1}^{M} g(\lambda_\alpha)
= \int_0^{\pi\over 2\eta} d\lambda \  \sigma(\lambda)\ g(\lambda) 
- {1\over N}\sum_{\alpha=1}^{\nu} g(\tilde\lambda_\alpha)
- {1\over 2N} \left[g(0) + g({\pi\over 2\eta}) \right]  
\label{euler} 
\ee
(plus terms that are of higher order in $1/N$), where $g(\lambda)$ is 
an arbitrary function, and $\{ \tilde\lambda_{\alpha} \}$ are the hole 
rapidities. 

Using the above result, we obtain the integral equation 
\be 
\sigma(\lambda) = 2 a_1(\lambda) 
+ {1\over N} \Big\{ a_1(\lambda) + b_1(\lambda) + a_2(\lambda) + b_2(\lambda)  
+ a_{2\xi_+ -1}(\lambda) + a_{2\xi_- -1}(\lambda) \non \\ 
+ a_2(\lambda - \tilde\lambda) + a_2(\lambda + \tilde\lambda) \Big\} 
- \int_0^{\pi\over 2\eta} d\lambda' \left[ a_2(\lambda - \lambda') 
+ a_2(\lambda + \lambda') \right] \sigma (\lambda')  \,, \qquad \lambda > 0
\ee 
where $\tilde \lambda$ is the hole rapidity, and
\be
a_{n}(\lambda) &=& {1\over 2\pi}{d \over d \lambda} q_{n}(\lambda) =
{\eta\over \pi} {\sinh (\eta n)\over \cosh(\eta n) - \cos (2 \eta 
\lambda)} \,, \\
b_{n}(\lambda) &=& {1\over 2\pi}{d \over d \lambda} r_{n}(\lambda) =
{\eta\over \pi} {\sinh (\eta n)\over \cosh(\eta n) + \cos (2 \eta 
\lambda)} = a_{n}(\lambda \pm {\pi\over 2 \eta})  \,.
\ee
Note that the $b_n(\lambda)$ terms are absent from the integral 
equation in the isotropic limit. The $b_1$ term originates 
from the factor $g_{1}$ in the BA Eqs. (\ref{compact}), and the
$b_2$ term originates from the last term in Eq. (\ref{euler}).

The symmetric density $\sigma_s(\lambda)$ defined by 
\be
\sigma_s(\lambda) = \left\{  
\begin{array}{lll} 
         \sigma(\lambda)  & \lambda > 0 \\
         \sigma(-\lambda) & \lambda < 0  
\end{array} \right. 
\label{sigma2}  
\ee
can now be readily found with the help of Fourier transforms, for 
which we use the following conventions:
\be
f(\lambda) = {\eta \over \pi} \sum_{k=-\infty}^\infty  e^{-2 i \eta k \lambda}
\hat f(k) \,, \qquad
\hat f(k) = \int_{-\pi/2\eta}^{\pi/2\eta}d\lambda\ e^{2 i \eta k \lambda}
f(\lambda) \,.
\label{fourier1}
\ee
Indeed, we find
\be
\sigma_{s}(\lambda) = 2 s(\lambda) + {1\over N} r^{(+)}(\lambda) \,, 
\label{sigma3}
\ee
where
\be 
r^{(+)}(\lambda) =  s(\lambda) + K(\lambda) + J(\lambda) + L(\lambda)
+ J_+^{(+)}(\lambda) + J_-^{(+)}(\lambda) + J(\lambda - \tilde\lambda) 
+ J(\lambda + \tilde\lambda) \,,
\label{rplus}
\ee 
and 
\be
\hat s = {\hat a_{1}\over 1 + \hat a_{2}} \,, \qquad
\hat J = {\hat a_{2}\over 1 + \hat a_{2}} \,, \qquad 
\hat J_{\pm}^{(+)} = {\hat a_{2\xi_{\pm}-1}\over 1 + \hat a_{2}} \,, \qquad 
\hat K = {\hat b_{1}\over 1 + \hat a_{2}} \,, \qquad 
\hat L = {\hat b_{2}\over 1 + \hat a_{2}} \,,
\label{fourier2}
\ee 
with
\be
\hat a_{n}(k) = e^{-\eta n |k|} \,, \qquad 
\hat b_{n}(k) = (-)^{k}\ \hat a_{n}(k) \,. 
\label{fourier3}
\ee 
Note that the Fourier series for $J_{\pm}^{(+)} (\lambda)$ converges 
for $\xi_\pm > 1/2$.

\section{Boundary $S$ Matrix}

The boundary $S$ matrix has the diagonal form
\be
{\cal K}(\tilde\lambda \,, \xi_\pm) = 
\left( \begin{array}{ll}
      \alpha(\tilde\lambda \,, \xi_\pm) &0  \\
      0  & \beta(\tilde\lambda \,, \xi_\pm)  \end{array}\right) 
\,. \label{form} 
\ee 
The matrix elements 
$\alpha(\tilde\lambda \,, \xi_\pm)$ and $\beta(\tilde\lambda \,, \xi_\pm)$ 
are the boundary scattering amplitudes for one-hole states
with $S^z = +{1\over 2}$ and $S^z = -{1\over 2}$, respectively.

We first compute $\alpha(\tilde\lambda \,, \xi_\pm)$. Setting 
\be
\alpha(\tilde \lambda \,, \xi_{\pm}) = 
e^{i \phi(\tilde \lambda\,, \xi_{\pm})} \,, 
\ee
we obtain by a calculation completely analogous to the one in 
Ref. \cite{DMN} the result
\be
\Phi^{(+)}(\tilde \lambda) \equiv
\phi(\tilde \lambda\,, \xi_{+}) + \phi(\tilde \lambda\,, \xi_{-}) =
2\pi\int_0^{\tilde\lambda} r^{(+)}(\lambda)\ d\lambda
+ const \,.   
\ee 
Recalling the result (\ref{rplus}) for $r^{(+)}(\lambda)$, 
and using the fact
\be
\int_0^{\tilde\lambda} \left[ J(\lambda - \tilde\lambda) 
+ J(\lambda + \tilde\lambda) \right] d\lambda\ 
= \int_0^{\tilde\lambda}  2 J ( 2\lambda )\ d\lambda\ \,,
\ee
we obtain
\be
\phi(\tilde \lambda\,, \xi_{\pm}) = \pi \int_0^{\tilde\lambda}  \left[ 
s(\lambda) + K(\lambda) + J(\lambda) + L(\lambda)
+ 2 J(2\lambda) + 2 J_{\pm}^{(+)}(\lambda) \right] d\lambda \,.
\ee 
We now use Eqs.  (\ref{fourier1}), (\ref{fourier2}), (\ref{fourier3}) 
to explicitly write the integrand as a Fourier series. Performing the
$\lambda$ integration, using the identity 
\be
\sum_{k=1}^{\infty} {e^{- 2 \eta k x}\over 1 + e^{- 2 \eta k}} 
 {1\over k}  = \log \left[ 
{\Gamma_{q^{4}}\left({x\over 2}\right)\over
\Gamma_{q^{4}} \left({x + 1\over 2}\right)} \right]
- {1\over 2} \log (1 - q^{4}) \,,
\ee
where $q = e^{-\eta}$ and $\Gamma_{q}(x)$ is the $q$-analogue of the 
Euler gamma function (see Appendix), and using also the $q$-analogue of the 
duplication formula \cite{GR}
\be
\Gamma_{q}(2 x)\ \Gamma_{q^{2}}({1\over 2}) =
(1 + q)^{2 x -1}\ \Gamma_{q^{2}}(x)\ \Gamma_{q^{2}}(x + {1\over 2})
\,, 
\ee
we obtain the following result for 
$\alpha(\tilde\lambda \,, \xi_\pm)$ (up to a rapidity-independent phase factor):
\be
\alpha(\tilde\lambda \,, \xi_\pm) = q^{-4 i \tilde\lambda}
{\Gamma_{q^8} \left({-i\tilde\lambda\over 2} + {1\over 4}\right) \over
  \Gamma_{q^8} \left({i\tilde\lambda\over 2} + {1\over 4}\right)} 
{\Gamma_{q^8} \left({i\tilde\lambda\over 2} + 1\right) \over
\Gamma_{q^8} \left({-i\tilde\lambda\over 2} + 1\right)}
{\Gamma_{q^4} \left({-i\tilde\lambda\over 2} + {1\over 4}(2\xi_\pm -1)\right)
\over 
  \Gamma_{q^4} \left({i\tilde\lambda\over 2} + {1\over 4}(2\xi_\pm -1)\right)} 
{\Gamma_{q^4} \left({i\tilde\lambda\over 2} + {1\over 4}(2\xi_\pm +1)\right)
\over
\Gamma_{q^4} \left({-i\tilde\lambda\over 2} + {1\over 4}(2\xi_\pm +1)\right)} 
\,. \label{result1} 
\ee 

We turn now to the computation of $\beta(\tilde\lambda \,, \xi_\pm)$, 
for which we must consider a one-hole state with $S^z = -{1\over 2}$.  
Instead of taking the pseudovacuum to be the ferromagnetic state with 
all spins up as we have done so far, we now take the pseudovacuum to 
be the ferromagnetic state with all spins down.  The expression 
(\ref{energy}) for the energy eigenvalues remains the same, the 
expression (\ref{spin}) for the $S^z$ eigenvalues becomes
\be
S^{z} =  M - {N\over 2}  \,,
\ee
and \cite{sklyanin} there is a change $\xi_\pm \rightarrow -\xi_\pm$ 
in the BA Eqs. (\ref{BA}).

We focus on the BA state consisting of one hole ($M = {N\over 2} - 
{1\over 2}$ with $\{ \lambda_\alpha \}$ real), which evidently has 
$S^z = -{1\over 2}$.  The corresponding function $r^{(-)}(\lambda)$ is 
the same as $r^{(+)}(\lambda)$ (see Eq.  (\ref{rplus})), except that now 
$J_{\pm}^{(+)}(\lambda)$ is replaced by $J_{\pm}^{(-)}(\lambda)$, with
Fourier transform
\be
\hat J_{\pm}^{(-)} = - {\hat a_{2\xi_{\pm}+1}\over 1 + \hat a_{2}} \,.
\ee

We observe that
\be
{\beta(\tilde\lambda \,, \xi_{-})\  \beta(\tilde\lambda \,, \xi_{+})\over
\alpha(\tilde\lambda \,, \xi_{-})\ \alpha(\tilde\lambda \,, \xi_{+})} = 
 e^{i 2 \pi\int_0^{\tilde\lambda} 
 \left[ r^{(-)}(\lambda) - r^{(+)}(\lambda) \right]\ d\lambda}
 \,.
\ee
Using the identity 
\be
J_{\pm}^{(-)}(\lambda) - J_{\pm}^{(+)}(\lambda) 
= - a_{2\xi_{\pm}-1}(\lambda) \,,
\ee
we conclude that 
\be
{\beta(\tilde\lambda\,, \xi_\pm)\over \alpha(\tilde\lambda\,, \xi_\pm)} =
- e_{2\xi_\pm-1}(\tilde\lambda) \,.
\label{result2}
\ee
We have fixed the sign in Eq. (\ref{result2}) by demanding that 
${\cal K}(\tilde\lambda \,, \xi_\pm)$
be proportional to the unit matrix for $\tilde\lambda = 0$.

\section{Discussion}

Our final result for the boundary $S$ matrix of the XXZ chain is
\be
{\cal K}(\tilde\lambda \,, \xi_\pm) = \alpha(\tilde\lambda \,, \xi_\pm)
\left( \begin{array}{lc}
      1  & 0  \\
      0  &  - e_{2\xi_\pm-1}(\tilde\lambda) \end{array}\right) 
\,,
\ee 
where $\alpha(\lambda\,, \xi_\pm)$ is given by Eq.  (\ref{result1}).  
It can be shown that this result agrees with the one found by Jimbo 
{\it et al.} \cite{jimbo} by means of the vertex operator approach.  
In the isotropic limit $\eta \rightarrow 0$, we see that $q 
\rightarrow 1$ and $\Gamma_{q}(x) \rightarrow \Gamma(x)$; hence, we 
recover the boundary $S$ matrix of the XXX chain \cite{GMN}, 
\cite{DMN}.

It would be interesting to see if this Bethe Ansatz method can be 
extended to the critical regime $|q|=1$, which is outside of the 
domain of the vertex operator approach.  

The $R$ matrix for the XXZ chain is associated with the fundamental 
representation of $A^{(1)}_{1}$. The present work opens the way to 
calculating boundary $S$ matrices for spin chains whose $R$ matrices
are associated with the fundamental representation of any affine Lie
algebra. For higher representations, the ground state involves complex
strings, and the analysis is more complicated. We hope to address 
these and related questions in the near future.

\section{Acknowledgments}

This work was supported in part by the National Science Foundation 
under Grant PHY-9507829.

\section{Appendix}

Here we prove the identity
\be
\sum_{k=1}^{\infty} {e^{- 2 \eta k x}\over 1 + e^{- 2 \eta k}} 
 {1\over k}  = \log \left[ 
{\Gamma_{q^{4}}\left({x\over 2}\right)\over
\Gamma_{q^{4}} \left({x + 1\over 2}\right)} \right]
- {1\over 2} \log (1 - q^{4}) \,,
\label{appendixresult}
\ee
where $q = e^{-\eta}$ and $\Gamma_{q}(x)$ is the $q$-analogue of the 
Euler gamma function, which is defined \cite{GR} as
\be
\Gamma_{q}(x) = (1 - q)^{1-x} \prod_{j=0}^{\infty}
\left[ {\left( 1-q^{1+j} \right)\over \left(1-q^{x+j} \right)} \right]
\,, \qquad 0 < q < 1 \,.
\ee

It is convenient to first consider the sum
\be
S(x) = \sum_{k=1}^{\infty} {e^{- 2 \eta k x}\over 1 + e^{- 2 \eta k}}
\,. \label{sum}
\ee
Expanding the denominator in an infinite series and then 
interchanging the order of summations, we obtain
\be
S(x) &=& \sum_{k=1}^{\infty} e^{- 2 \eta k x} 
\sum_{n=0}^{\infty} (-)^{n} e^{- 2 \eta k n} \\
&=& \sum_{n=0}^{\infty} (-)^{n} \sum_{k=1}^{\infty} 
e^{- 2 \eta k (x + n)} \\
&=& \sum_{m=0}^{\infty} \left\{ 
{e^{- 2 \eta (x + 2 m)}\over 1 - e^{- 2 \eta (x + 2 m)}}  
- {e^{- 2 \eta (x + 2 m + 1)}\over 1 - e^{- 2 \eta (x + 2 m + 1)}}  
\right\} \\
&=& {1\over \log q^{4}} \left[ 
\psi_{q^{4}}\left({x\over 2}\right)
- \psi_{q^{4}}\left({x + 1\over 2}\right) \right] \,,
\label{sresult}
\ee
where 
\be
\psi_{q}(x) &=& {d\over dx} \log  \Gamma_{q}(x)  \\
&=& - \log (1 - q) + \log q \sum_{j=0}^{\infty}
{q^{x + j}\over 1 - q^{x + j}} \,.
\ee
Integrating the result (\ref{sresult}) with respect to $x$, and evaluating 
the integration constant by considering the limit $x \rightarrow 
\infty$, we obtain the desired identity Eq.  (\ref{appendixresult}).

\vfill\eject


\begin{thebibliography}{99}

\bibitem{GZ}
S. Ghoshal and A. B. Zamolodchikov, Int. J. Mod. Phys. {\it A9} (1994)
3841; {\it A9} (1994) 4353.

\bibitem{GMN}
M. Grisaru, L. Mezincescu and R.I. Nepomechie, J. Phys. {\it A28} (1995) 1027.

\bibitem{korepin}
V.E. Korepin, Theor. Math. Phys. {\it 76} (1980) 165;
V.E. Korepin, G. Izergin and N.M. Bogoliubov, {\it Quantum Inverse 
Scattering Method, Correlation Functions and Algebraic Bethe Ansatz}
(Cambridge University Press, 1993).

\bibitem{andrei/destri}
N. Andrei and C. Destri, Nucl. Phys. {\it B231} (1984) 445.

\bibitem{FS}
P. Fendley and H. Saleur, Nucl. Phys. {\it B428} (1994) 681.

\bibitem{essler}
F.H.L. E{\ss}ler, J. Phys. {\it A29} (1996) 6183.

\bibitem{asakawa/suzuki}
H.  Asakawa and M.  Suzuki, J.  Phys.  {\it A30} (1997) 3741; 
``Boundary scattering matrices in the Hubbard model with boundary 
fields'', University of Tokyo preprint.

\bibitem{tsuchiya}
O. Tsuchiya, J. Phys. {\it A30} (1997) L245.

\bibitem{DMN}
A.  Doikou, L.  Mezincescu and R.I. Nepomechie, J. Phys. {\it A30}
(1997) L507.

\bibitem{JM}
J.D. Johnson and B.M. McCoy, Phys. Rev. {\it A6} (1972) 1613.

\bibitem{GR}
G.  Gasper and M.  Rahman, {\it Basic Hypergeometric Series} 
(Cambridge University Press, 1990).

\bibitem{jimbo}
M. Jimbo, R. Kedem, T. Kojima, H. Konno and T. Miwa,
Nucl. Phys. {\it B441} (1995) 437.

\bibitem{gaudin1}
M. Gaudin, Phys. Rev. {\it A4} (1971) 386; 
{\it La fonction d'onde de Bethe} (Masson, 1983).

\bibitem{alcaraz}
F.C. Alcaraz, M.N. Barber, M.T. Batchelor, R.J. Baxter and G.R.W. Quispel,
J. Phys. {\it A20} (1987) 6397. 

\bibitem{sklyanin}
E.K. Sklyanin, J. Phys. {\it A21} (1988) 2375.

\bibitem{FT}
L.D. Faddeev and L.A. Takhtajan, J. Sov. Math. {\it 24} (1984) 241.


\end{thebibliography}
\end{document}